\documentclass[journal]{vgtc}                     

\onlineid{1010}

\vgtccategory{Research}

\vgtcpapertype{Technique}

\title{Visualizing Spatial Semantics of Dimensionally Reduced Text Embeddings}

\author{%
  Wei Liu,
  Chris North, and 
  Rebecca Faust
}

\authorfooter{
  \item
        Wei Liu, Computer Science, Virginia Tech, Blacksburg, Virginia, United States.
  	E-mail: wliu3@vt.edu.
  \item
  	Chris North, Sanghani Center for AI and Data Analytics, Virginia Tech, Blacksburg, Virginia, United States.
  	E-mail: north@cs.vt.edu.

  \item Rebecca Faust, Computer Science, Tulane University, New Orleans, Louisiana, United States.
  	E-mail: rfaust1@tulane.edu.
}

\abstract{
Dimension reduction (DR) can transform high-dimensional text embeddings into a 2D visual projection facilitating the exploration of document similarities. However, the projection often lacks connection to the text semantics, due to the opaque nature of text embeddings and non-linear dimension reductions. To address these problems, we propose a gradient-based method for visualizing the spatial semantics of dimensionally reduced text embeddings. This method employs gradients to assess the sensitivity of the projected documents with respect to the underlying words. The method can be applied to existing DR algorithms and text embedding models. Using these gradients, we designed a visualization system that incorporates spatial word clouds into the document projection space to illustrate the impactful text features. We further present three usage scenarios that demonstrate the practical applications of our system to facilitate the discovery and interpretation of underlying semantics in text projections.

}


\teaser{
  \centering
  \includegraphics[width=\linewidth, alt={}]{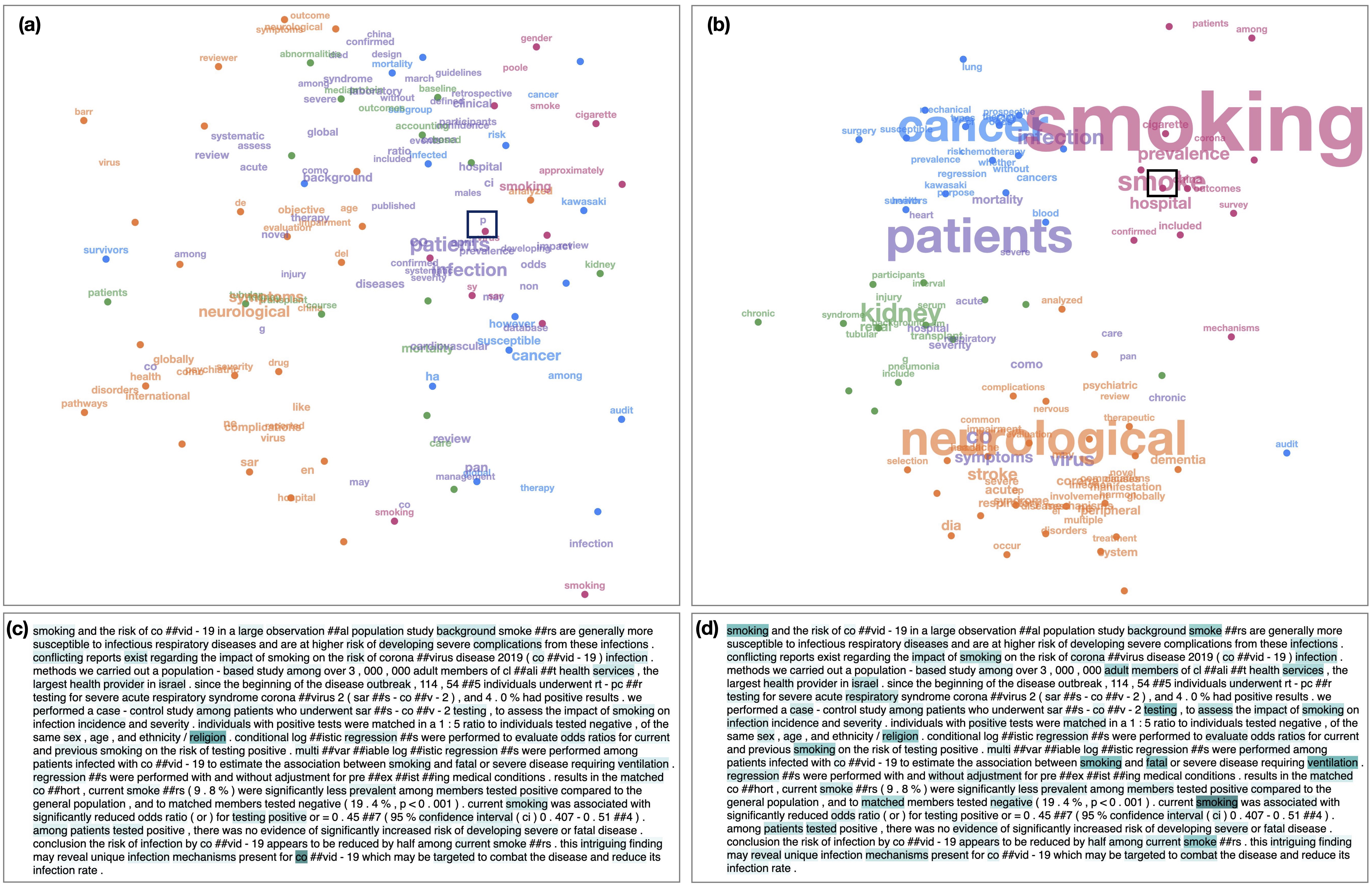}
  \caption {Document projections of COVID-19 open research articles with gradient-based word explanations. (a) A projection from a pre-trained BERT model, featuring a spatial word cloud showing key words that impact the document projection. (b) Projection from a fine-tuned BERT model, refined based on the data domain. The resulting spatial word cloud captures and explains the differences between the two embedding spaces, and how well the refined model captures the domain.   (c) The projection heatmap for a selected document, illustrating the individual word impact on document 2D positioning. (d) Refined projection heatmap of the same document, highlighting the word "smoking", which reflects the domain. }
  \label{fig:case_study_covid19}
}

\graphicspath{{figs/}{figures/}{pictures/}{images/}{./}} 

\usepackage{tabu}                      
\usepackage{booktabs}                  
\usepackage{lipsum}                    
\usepackage{mwe}                       

\usepackage{mathptmx}                  
\usepackage{amsmath}   
\usepackage{algorithm}
\usepackage{algpseudocode}

\begin{document}


\firstsection{Introduction}
\maketitle

Dimension Reduction (DR) techniques are popular for visualizing and analyzing collections of text documents. They enable users to explore patterns, trends, and relationships within the data in a low-dimensional space, which is more intuitive for human understanding \cite{rudin2022interpretable}.  Recent advancements in deep learning models have significantly improved our capacity to process complex textual data. The text embeddings extracted from these models capture the contextual relationships between documents, enabling the projection of documents into a two-dimensional space for exploratory analysis~\cite{bian2021semantic}. The spatial layout in the projection space naturally aligns with human cognitive processes, through the visual \emph{proximity} $\approx$ \emph{similarity} metaphor\cite{bian2021semantic, marshall1994viki}. Specifically, the spatial distance of documents in the projection represents their relative similarity, providing insights into document relationships. 

However, interpreting the projections of text embeddings generated by DR techniques remains a challenge\cite{rudin2022interpretable}. They lack an inherent connection to the semantics of the underlying text. Users often struggle with questions such as "Why is this document positioned here", or "What causes these documents to form a cluster in the projection space?". In the absence of methods to help answer these questions, users often must inspect individual documents to reason about similarities, somewhat defeating the purpose of DR in the first place.  Thus, providing insights into these questions is important for the exploratory analysis of DR projections. 

Recent efforts explored gradient-based explanations for DR projections. Faust et al. proposed DimReader to use gradients to explain non-linear projections for structured datasets \cite{faust2018dimreader}. By calculating the gradients of the projection with respect to underlying features, DimReader generated axis lines that reveal the sensitivities of the projection to underlying data features\cite{faust2018dimreader}. This method shows the impact of input data features on the DR space, helping users understand the complex projection space. However, DimReader is designed for structured datasets, where the high-dimensional data features are clearly defined and interpretable. In contrast, text data is first embedded into high-dimensional representations, via deep learning embedding models, that lack interpretable features and are then projected, making analysis and understanding of the DR space more complicated. Therefore, a DRs of text embeddings require specialized approaches to capture and illustrate the semantics of the DR space.

To address these problems, we propose a system to visualize the spatial semantics of text embedding DRs to enhance the understanding of document projections, providing users with intuitive and informative visualizations that assist in reasoning about document placement and clustering.  This system leverages projection gradients derived from the embedding model to illustrate the influence of underlying text features on document positioning within the 2D projection spaces. 
Specifically, we calculate the projection gradients with respect to individual words to capture their impacts on  the document in the projection space. Additionally, we introduce spatial word clouds to display the words that most significantly influence documents, aggregating common words across documents, while respecting the spatial layout of the projection space.
Words are overlayed within the projection space to reflect the layout of the documents they originate from, enhancing the understanding of document organization patterns.

The contributions of this paper include:

\begin{itemize}
    \item  A method for calculating gradients in DRs of text embeddings
  
    \item  A visualization system that integrates gradient impacts into the projection space via spatial word clouds
   
    \item Three usage scenarios to demonstrate the practical applications of our system in data analysis tasks, showing how they uncover meaningful insights from text projections.
\end{itemize}

\section{Related Work}

\subsection{Visualizing Text Corpora with Projections}
DR techniques have been widely applied to text data visualization and analysis, allowing us to transform complex, high-dimensional data into more manageable forms\cite{liu2018bridging}.  Researchers have developed interactive visual analytics systems using DR to enhance text analytics. Typograph, for example, offers a multi-scale detailed view of documents within a single spatial projection \cite{endert2013typograph}. 
Endert et al. further developed ForceSPIRE, a system that supports semantic interaction for text visual analytics, allowing users to interact with the system based on their interests and domain knowledge. The underlying model learns from user interactions, updating the projected layout accordingly \cite{endert2012semantic, endert2011unifying}. Building upon this, StarSPIRE incorporates multi-model semantic interaction techniques\cite{bradel2014multi, bradel2015big}, while SIRIUS\cite{dowling2018sirius} provides a dual, symmetric projection for both attributes and observations, and Cosmos facilitates sensemaking with large text corpora\cite{dowling2019interactive}. Recent work such as DeepSI integrates deep learning with the human sensemaking loop, allowing for more efficient semantic interaction inference \cite{bian2021deepsi}. However, a limitation of DeepSI and other methods discussed above is their limited illustration of the spatial semantics of the DR space. Thus, our work aims to capture and present the semantics of the text projection space.

\begin{figure*}[t]
    \centering
    \vspace{-1em}
    \includegraphics[width=.9\linewidth]{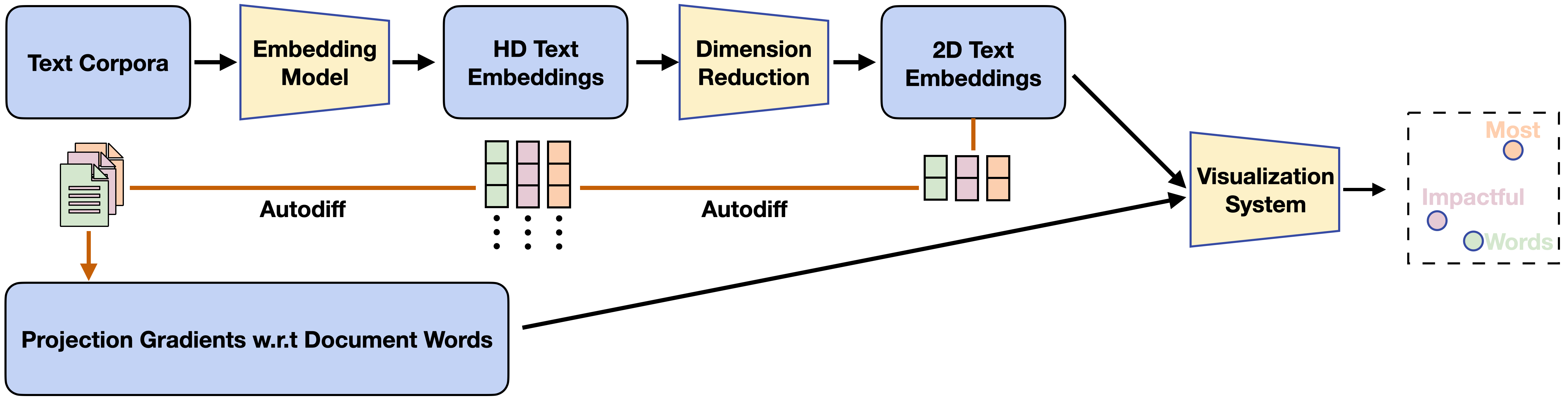}
    \vspace{-.5em}
    \caption{The pipeline for our system. In a forward pass, we first embed the documents into an high-dimensional (HD) space and then we project them to two dimensions with DR. Next, we perform a backward pass using Autodiff through the pipeline that calculates the gradients of the 2D embeddings with respect to the document words. Finally, the 2D embeddings and gradients are passed to the visualization system to create the visualizations.}
    \vspace{-1.5em}
    \label{fig:text_pipeline}
\end{figure*}

\subsection{Recovering the Semantics of Embeddings}

Counterfactual explanation, popular for improving the interpretation of the machine learning model \cite{wang2024empirical}, describes how input features must change to alter the model's outcome. This method has been applied by tools such as What-IF Tool \cite{wexler2019if}, ViCE \cite{gomez2020vice}, AdViCE \cite{gomez2021advice}, DECE \cite{cheng2020dece}, and Interact \cite{ciorna2023interact}. By employing counterfactual explanations for DR projections, Bian et al. introduced DeepSE to provide semantic explanations of DR \cite{bian2021semantic}. This approach involves generating a counterfactual projection by removing one feature (e.g. instances of a word) each time from a document to identify and display the most impactful features. However, this technique, by removing all instances of the word, could fundamentally alter the document's semantic meaning. Moreover, the semantic meaning of the same word can vary based on its context, introducing additional complexities.

Feature contribution explanations are another popular approach. It investigates how the features contribute to the model's outcome. Tools such as LIME \cite{ribeiro2016should}, SHAP \cite{lundberg2017unified}, and Integrated Gradients \cite{sundararajan2017axiomatic} have been widely applied to classification and prediction tasks. These methods have been adapted to provide explanation in the DR context. Zang et al. introduced DMT-EV, an explainable deep neural network approach for DR, integrated within an interactive visual interface \cite{zang2022evnet}. This system supports explanations for feature contributions at global, local, and transformation levels, and aids in community discovery \cite{zang2022evnet}. Additionally, Marcilio-Jr et al. present ClusterShapley, a visual analytic system that offers the explanation of clusters in DR projections using Shapley values, and highlights the feature contribution to cluster formation \cite{marcilio2021explaining}. However, the research of understanding DR in text data remains underexplored, given that the text embedding is abstract and lacks interpretable features.

Several existing works use gradient-based explanations to enhance interpretability in machine learning. By analyzing the gradients of the model output with respect to the input feature, these methods quantify the features' contribution to the model's output. Techniques such as Integrated Gradients \cite{sundararajan2017axiomatic}, and Grad-CAM \cite{selvaraju2017grad} are commonly used in the image explanation, emphasizing the important image features for classification and prediction tasks \cite{panwar2020deep, moujahid2022combining, vcik2021explaining}. Similarly, these approaches have been extended to text data, highlighting the keywords that significantly influence the model output \cite{kokhlikyan2020captum, gorski2020towards}. Recently, DimReader used gradients to illustrate the impact of input features from the original dataset on the projection space. It incorporates generalized axes to visually represent these effects\cite{faust2018dimreader}. However, DimReader and similar gradient-based explanation methods for DR primarily focus on data with interpretable features \cite{hamal2023interpreting}. In this paper, we expand on existing approaches to create gradient-based explanations for DRs of text embeddings, aiming to visualize and enhance understanding of text projection spaces. This addresses the critical need for explainability in the analysis of complex text data in DR.

\subsection{Visualizing Semantics of Text}
Many popular approaches for visualizing the semantics of text use word clouds to illustrate key topics. These have been used to create spatial illustrations of text semantics in individual documents~\cite{viegas2009participatory, 4577920}, time-varying text data~\cite{chi2015morphable}, contrastive analysis of multiple texts~\cite{lohmann2015concentri, john2018multicloud}.
Some of these methods aim to spatially organize the words in the cloud by topic~\cite{duarte2014nmap, hearst2019evaluation, xu2016semantic, burch2014radcloud, wang2020recloud}. Paulovich et al. presented ProjCloud, a method for creating word clouds in DRs based on the document keywords in clusters of documents~\cite{paulovich2012semantic}. Our system adopts a word-cloud based approach to present the impactful words in a projection while grounding them in the spatial organization of the DR. 
While our approach is similar to ProjCloud, ProjCloud, like other keyword approaches, will not capture the semantics of the DR space itself, but rather the semantics of groups of documents analyzed independent of this space.

\section{Methodology}
\label{sec:methodology}

In this section, we describe our methodology for projecting documents and calculating their projection gradients. Figure~\ref{fig:text_pipeline} presents an overview of this process. 

\subsection{Text Embedding and Projection}
Our method begins by processing the text documents with an embedding model, as seen in the pipeline in \cref{fig:text_pipeline}. For our examples, we will use the BERT (Bidirectional Encoder Representations from Transformers) \cite{devlin2018bert} model but, as is discussed below, any embedding model compatible with popular deep learning libraries can be substituted. The embedding model transforms the documents into high-dimensional (HD) embedding vectors. These embeddings preserve the contextual relationships within the corpus, thereby enabling a meaningful projection through DR. Following this, we use DR to project the embeddings into low-dimensional space. For examples in this paper, we use MDS and t-SNE but other DRs can easily be substituted.

\subsection{Calculating Gradients of Text Embedding Projections}
To effectively illustrate the semantics of the projection space, we must identify the the text features that most influence this space. In past work, DimReader demonstrated how gradients of projected coordinates with respect to data features offer us insight into how these features impact the projection space, enabling people to more effectively interpret the generated DR space ~\cite{faust2018dimreader}.  However, this approach assumes that the data input into the DR has interpretable data features (e.g. named columns in a table).  For projections of text embeddings, the data features passed into the DR are abstract features generated by an embedding model with no human-interpretable meaning. Though these embeddings encode the semantics of documents and capture contextual relationships between documents, they do not provide insight into the text features they encode. Thus, to recover this context, we must calculate the gradients through both the projection and the embedding mode, back down to the underlying text features (i.e. individual words). 

\subsubsection{Notation}
To calculate the impact of individual words on the projected location, we calculate the gradients of the projected coordinates with respect to the individual words in the corresponding document. We then organize these into a structure called a tangent map $M$, such that for a document (organized as an ordered list of words) $p_i$ the gradients of the $x$ and $y$ projected coordinates are given by

\begin{equation}
    M_{i} = \left[ \begin{array}{ccc}
    \frac{\partial v_{i,x}}{\partial p_{i,f_1}} & \dots & \frac{\partial v_{i,x}}{\partial p_{i,f_d}} \\
    \frac{\partial v_{i,y}}{\partial p_{i,f_1}} & \dots & \frac{\partial v_{i,y}}{\partial p_{i,f_d}}
    \end{array} \right]
    \label{eq:grad_define}
\end{equation}

where $P_i=\{p_{i,f_1}\dots p_{i,f_d}\}$ is the document with $d$ words $f_1 \dots f_d$ and $V_i =\{v_{i,x}, v_{i,y}\}$ is the low-dimensional projected representation.

Thus, for a document containing $d$ words, the impact of the $j$-th word on the projection point $v_i$ in the 2D space is given by the partial derivatives of the coordinates of $v_i$ with respect to $f_j$. This partial derivative, captured in the gradient of $v_i$, is denoted as $M_{i, f_j}$ and given by:

\begin{equation}
    M_{i, f_j} = \left[ \frac{\partial v_{i,x}}{\partial f_{j}}, \frac{\partial v_{i,y}}{\partial f_{j}} \right]
     \label{eq:grad_token}
\end{equation}

The partial derivative, as defined in \cref{eq:grad_token}, can be considered as a vector, where its direction within the project space indicates where the word tends to "pull" the projected point. The magnitude of this vector represents the word's relative impact on the document's positioning within the 2D projection space and is calculated as follows:

\begin{equation}
    \text{Magnitude} = \| M_{i, f_j} \| = \sqrt{\left(\frac{\partial v_{i,x}}{\partial f_{j}}\right)^2 + \left(\frac{\partial v_{i,y}}{\partial f_{j}}\right)^2}
    \label{eq:grad_token_mag}
\end{equation}

This approach allows us to evaluate the impact of individual words on the spatial arrangement of documents in the projection, enhancing understanding of the semantic relationships in the data. 

Note, for simplicity, in the remainder of this paper, we refer to the partial derivatives of the projected points with respect to a given feature (i.e.~\cref{eq:grad_token}) as the ``gradient impact'' of that feature.

\subsubsection{Calculating Projection Gradients of Text Embeddings}
In past work, DimReader demonstrated a method for calculating the gradients of projected coordinates, with respect to underlying data features. They employed a method called automatic differentiation (autodiff) to calculate the partial derivatives of the projected coordinates with respect to each feature, during the execution of a DR, generating the gradient \cite{griewank2008evaluating} and compiled them into a tangent map (see \cref{eq:grad_define}). 

DimReader uses “forward mode” autodiff to automate gradient calculations as the algorithm executes \cite{faust2018dimreader}. This approach employs an extended that captures both the value and derivative of the current variable, with respect to a specified value.  Each variable $x$ is replaced by a dual number $x = (a,b)$, where $a$ represents the actual value of $x$ and $b$ represents the partial derivative of $x$ with respect to a specific variable of interest. As the DR calculates the projected coordinates, it simultaneously calculates the specified partial derivative. However, DimReader implements this number system through operator overloading, which many existing methods do not allow as input. This limits DimReader's generalizability and compatibility with existing methods. Additionally, the extended number system only calculates partial derivatives with respect to a single variable in each execution. This means that, to generate the gradients for each point, we must execute the DR $nd$ times where $n$ is the number of points in the dataset, and $d$ is the number of high-dimensional features. In large datasets and complex algorithms, this becomes prohibitively expensive. This inhibits the applicability of DimReader to text embeddings as (1) dual numbers are not compatible with the deep learning libraries used to build embedding models and (2) even if it were compatible, documents are very high dimensional (i.e. they have many words) making DimReader prohibitively expensive for text embedding projections.  Thus, we must overcome these limitations to capture gradients in text embedding projections. 

To do so, our method uses a different autodiff mode - “backward mode”, which constructs a computation graph that traces all operations from a given starting point, in this case the original data point ($p_i$), to a final value, i.e., the projected coordinates ($v_i$), thereby enabling backward propagation to calculate the gradient $[\frac{\partial v_i}{\partial p_{i,0}}, ... ,\frac{\partial v_i}{\partial p_{i,d}}]$. "Backward mode" has two key advantages. Firstly, it allows us to compute the gradients that define $M_i$ in just 2 passes (one for $v_{x}$ and one for $v_{y}$). Therefore, for a document with $d$ features, this results in $2d$ backward passes in "backward mode", compared with $dn$ passes (one forward pass per feature per point) required in the “forward mode”. Secondly and importantly, “backward-mode” autodiff is supported by common deep learning libraries, such as the Torch \cite{collobert2011torch7} (via autograd) and Tensorflow \cite{abadi2019tensorflow}. Thus, we enable gradient calculations in embedding models built in Torch. 

By leveraging the built-in capabilities of deep learning libraries, we enable the calculation of gradients through the embedding model. Now, to capture the gradients throughout the entire process (from words to projected coordinates), we must also adapt our projection methods to be compatible with Torch tensors. For most methods, this simply means creating an implementation compatible with library, e.g. compatible with Torch tensors.  For example, for MDS, we extracted the MDS algorithm from the Sklearn library \cite{pedregosa2011scikit} and made minor modifications to make it compatible with Torch. SImilarly, for t-SNE, we simply adapated the implementiation published by Van der Maaten~\cite{maaten-tsne-python}.  This allows us to use the built-in autodiff capabilities of Torch to calculate the tangent map of the projection through the entire pipeline, back down to the individual words in the document.

\section{Visualization Design}
This section introduces the visualization design of our system. Though the gradient calculation gives us a vector indicating the direction a word pulls the document, we found the directions to be less consistent and instance dependent. Thus, we opt to use the magnitude of the gradients to explain the projection of documents. We have the following design goals: 

\textbf{DG1: Detailed Word Impact Analysis for Each Document}: Enable users to learn how individual words impact the positioning of a document within the 2D projection space. This supports a detailed, document-by-document exploration of word significance. 

\textbf {DG2: Overview of the Most Impactful Words for Each Document within the Projection Space}. Provide an immediate, clear visualization of the words that most significantly influence the positioning of each document in the projection space. It offers a direct understanding of what influences document position, enabling users to discover patterns and identify connections.

\textbf {DG3: Aggregated Words to Enhance Spatial Understanding of Document Clusters}. Using spatial analysis to aggregate words, this goal allows users to quickly identify the key words that impact the projection location of documents. It also provides a comprehensive overview of document clustering and the significant words associated with the entire dataset affecting the document projection. The placement of aggregated words in the projection space correlated with the locations of projected documents, aiming to enhance user comprehension of document clustering patterns.

To address these goals, our system presents three visual designs: impact heatmaps, projections overlayed with impactful words for individual documents, and projections overlayed with spatial word clouds.  In the following, we describe each visualization in greater detail.

\subsection{Impact Heatmap}
To address DG1, our system creates an impact heatmap. 
When a user selects a document by clicking the document dot in the projection space, the system shows a heatmap representation of the document. This heatmap, as seen in \cref{fig:case_study_covid19} (c) and (d), visualizes the magnitude of each word's impact and provides direct contextual insights. 
The color intensity of each word's background correlates with its gradient magnitude. This provides a visual indicator of how much each word impacts the document's location in the projection space. 
The words themselves offer contextual information about the document, thus by enabling the user to quickly identify the most impactful words, the heatmap helps users understand what impacts the document's position.

\subsection{Impactful Words for Individual Documents}
For a concise overview of the space, our system overlays impactful words (addressing DG2) for individual documents on the projection to display the most significant words that influence the document positioning within the projection space. Each document in the projection space is associated with a marker representing its most impactful word. The size of the marker indicates the relative magnitude of the word impact. This allows people to quickly identify the most impactful word for individual documents, as well as look for patterns among similar documents (those placed in close proximity).

\subsection{Spatial Word Clouds}

\begin{figure*}
    \centering
    \vspace{-1em}
    \includegraphics[width=0.85\linewidth]{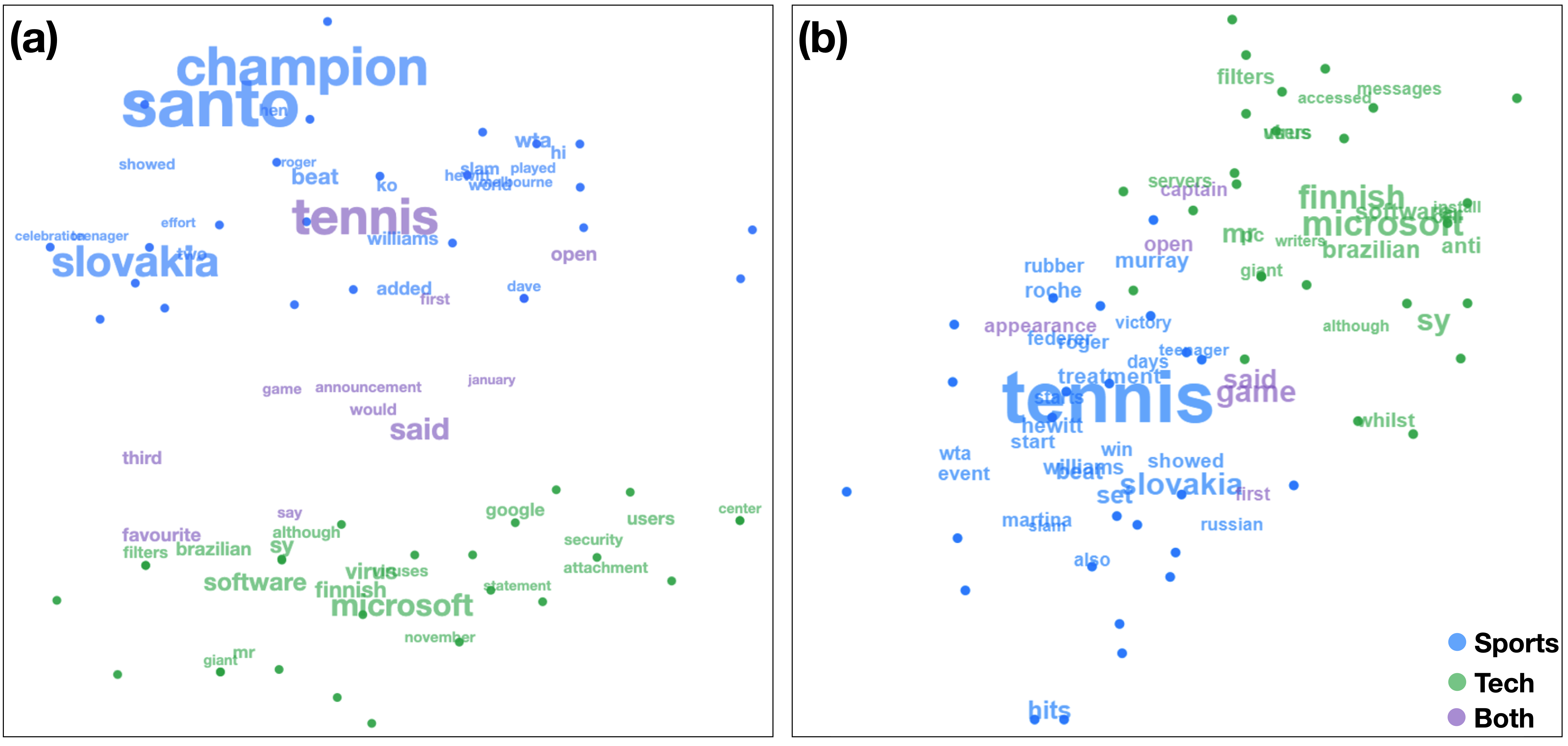}
    \vspace{-.5em}
    \caption{Comparing DR algorithms on a collection of news articles about sports and tech. (a) shows the spatial word cloud for the DR space generated by MDS. (b) shows the spatial word cloud for the DR space generated by t-SNE. We see that t-SNE identifies a strong central topic for the sports articles (``tennis''). MDS still picks up on this central topic but shows increased focus on subtopics within the sports articles, e.g. ``champion''}
    \vspace{-1.5em}
    \label{fig:case_study_bbc}
\end{figure*}

To enhance the understanding of how words impact the document's positioning within the projection space, we introduce spatial word clouds (addressing DG3). Spatial word clouds use spatial analysis to create word clouds where word positions are constrained by the spatial organization of the projected documents. This provides an intuitive visual representation to display words that significantly influence document placement. The word clouds aggregate words that appear in multiple documents to reduce visual clutter from label overlap. It strategically places aggregated words in the projection space to align with the spatial distribution of documents impacted by that word, facilitating visualization of document clustering trends. The key, high-level steps of this process are outlined below, with \cref{alg:spatial_tag_clouds} providing more detail.

\textbf { Step 1. Group Documents by Impactful Words}: First, we group documents based on their most impactful words (i.e. by default, words with top twenty gradient impacts), and form word groups of documents containing that word.

\textbf{Step 2. Centroid Calculation}: For each word group, we determine the centroid to denote the spatial location of the aggregated words. This calculation applies a weighted approach, where the frequency of a word's occurrence in each document influences its contribution to the centroids' location. For example, if a word appears $m$ times in document A, and $n$ times in document B within a cluster, then the centroid's position, $C$, is calculated to reflect these occurrences proportionately. This approach provides a representative spatial summary of the global word distribution.

\setlength{\textfloatsep}{5pt}
{\begin{algorithm}[t]
\caption{Generate Spatial Word Clouds.}
\label{alg:spatial_tag_clouds}
\begin{algorithmic}[1]
\Statex \textbf{Input:} Documents $D$, WordImpactScores $T$ 
\Statex \textbf{Output:} SpatialWordClouds
\State Initialize WordGroups as an empty map
\For{each document $d$ in $D$}
    \State impactfulWords $\gets$ IdentifyImpactfulWords($d$, $T$) 
    \For{each word $t$ in impactfulWords}
        \If{$t$ not in WordGroups}
            \State Initialize WordGroups[$t$] as an empty list
        \EndIf
        \State Append $d$ to WordGroups[$t$]
    \EndFor
\EndFor
\State Initialize WordClouds as an empty list
\For{each groupKey in WordGroups.keys()}
    \State $G \gets$ WordGroups[groupKey]
        \State centroid $\gets$ CalculateCentroid($G$)
        \Comment{Centroid Calculation: $C_x = (\Sigma_{x\in G}$ $ x_i / n)$, $C_y = (\Sigma_{x\in G}$ $  y_i / n)$}
        \State wordCloud $\gets$ Create a new WordCloud for $G$ with centroid
        \State Append wordCloud to WordClouds
\EndFor
\State \Return WordClouds
\end{algorithmic}
\end{algorithm}
}

Note, currently we use spatial word clouds to illustrate the global distribution of words, such that each word only appears once in the visualization. However, an additional step could be added between 1 and 2 to spatially subdivide word groups based on proximity, displaying the word at the centroid for each sub-group.

\textbf{Visualization of Word Clouds}: As seen in \cref{fig:case_study_covid19}, the size of each word within the cloud corresponds to its aggregated gradient magnitude. For documents with predefined labels, the words are colored to match the class of documents they impact, provided that all impacted documents have the same label. Otherwise, they are colored purple. For documents without predefined labels, we can first compute labels, and then color the words similarly. This involves using a clustering algorithm, such as KNN, to assign labels based on the document projection.
To maintain clarity, after grouping documents in Step 1, we filter out word instances that appear only once within a specific document. These instances might either be unique to a single document and not appear in others or present in several documents but not form part of the aggregated clusters. Furthermore, when multiple words share the same centroid-often because they present multiple times within a single document but don't aggregate with nearby documents-we display only the word with the highest gradient impact. This strategy helps enhance readability and also ensures that visualization highlights the most impactful information.
Importantly, the spatial placement of words correlates with the location of related documents within the 2D space, offering a direct visual mapping between word significance and document layout. 

The visualization design of our system incorporates impact heatmap, impactful words of individual documents, and Spatial Word Clouds to offer visual tools for interpreting document projections. It provides multiple layers of insights, from a detailed impact analysis of individual words on a single document to a broader analysis across the document collections, which facilitates a deeper understanding of textual data.

\section{Usage Scenarios}
In this section, we present three usage scenarios that illustrate how our visual system enhances understanding of document projections.

\begin{figure*}[t]
  \centering
   \vspace{-.5em}
  \includegraphics[width = 0.85\linewidth]{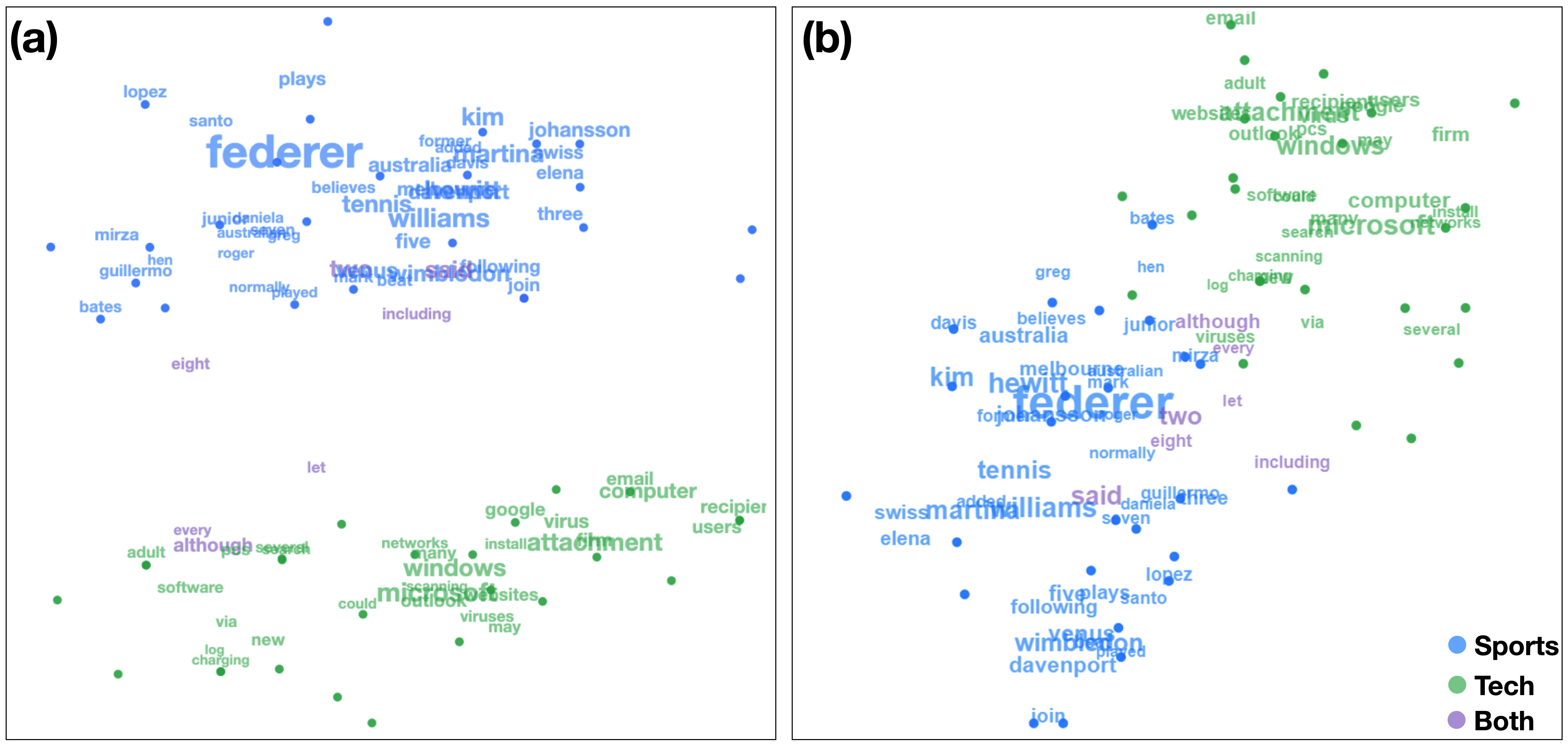}
  \vspace{-.5em}
  \caption {Spatial word clouds generated with attention values.  (a) shows the attention-based cloud for the DR space generated by MDS. (b) shows the attention-based cloud for the DR space generated by t-SNE. Unlike the gradient-based clouds in \cref{fig:case_study_bbc}, the impactful words identified by attention remain constant between DR algorithms, failing to explain the impact of the DR on the space.}
  \vspace{-1.5em}
  \label{fig:case_study_attention}
\end{figure*}

\subsection{Comparing Embedding Models}
\label{sec:covid_case_study}
In this first scenario, we illustrate the application of our system to compare the embedding spaces of two different embedding models: a pre-trained BERT model and a fine-tuned BERT model, refined to capture the data domain. We use the dataset of COVID-19 open research articles. This dataset categorizes documents into four different risk factors: cancer, chronic kidney disease, smoking status, and neurological disorders.  This example shows how our system illustrates the differences between different embedding spaces, projected with the same DR algorithm.

\cref{{fig:case_study_covid19}} (a) shows the spatial word clouds generated for the projection of the per-trained embedding space. Because we have labels for each document, the risk factor, each point  is colored by the risk factor (cancer: blue, chronic kidney disease: green, neurological disorders: orange, and smoking status: pink). Additionally, the impactful words are colored by the risk factor of the documents they impact, provided that all occurrences come from documents of the same risk factor. Words that impact documents of multiple risk factors are colored purple. We immediately notice that the pre-trained embedding space does not effectively capture the four underlying categories of documents. The  projection lacks a clear separation of the risk factors, which is emphasized by the lack of any pattern among the impactful words. Despite some medical-related words identified on the projection, there isn't a clear pattern relative to the risk factors. 

In contrast, \cref{{fig:case_study_covid19}} (b) uses the pre-trained BERT model fine-tuned to separate these four categories.  Now, the projection forms four clusters, each corresponding to a specific risk factor. The spatial word clouds illustrate the semantics of the refined projection space, highlighting the most impactful words that reflect the topic of each category. The words 'smoking", "neurological", "kidney", and "cancer" directly correspond to document categories, and have a significant impact on the current projection of documents. The placement of these words is consistent with the layout of corresponding documents in the projection space. We also see a set of general medical words shared among documents from different topics, such as "patients", "virus", "mortality", and "symptoms". Additionally, the larger size of words in the refined embedding space indicates a higher aggregation level of instances for these highlighted words, suggesting their frequent occurrence across documents and significant influence on clustering within the projection space. 

The impact heatmap offers further insights into the differences between the embeddings. In the pre-trained embedding, as seen in \cref{{fig:case_study_covid19}} (c), the words "religion" and "COVID" are the most impactful to the projection space for the document outlined in the black box. In the fine-tuned embedding space, the highlighted words for that specific document change to align with the risk factor, including "smoking", and "smoke". It indicates that the words identified by gradient methods successfully capture the data's domain.  

\subsection{Comparing DR Algorithms}
\label{sec:bbc_case_study}
In this user scenario, we illustrate the application of our system to compare the projection spaces generated by two different DR algorithms, MDS and t-SNE. In this example, we use a dataset of BBC news articles containing articles about sports and technology \cite{greene2006practical}. 
\cref{fig:case_study_bbc} (a) shows the spatial word cloud for the space generated by MDS while \cref{fig:case_study_bbc} (b) shows the spatial word cloud for t-SNE.  We see that both algorithms identify the same central topics - tennis for sports and Microsoft for tech. However, we see that t-SNE more strongly identifies "tennis" as a key topic in the dataset's sports news, as seen by the larger size of "tennis" than other words in the word cloud. In contrast, MDS captures more local features within the sports articles, demonstrated by the heavy impact of words such as ``slovakia'' and ``champion''. Thus, gradient-based cloud demonstrates the differences in the information prioritized by different DR algorithms.

\subsection{Gradient vs Attention-based Spatial Word Clouds}
Our approach goes beyond traditional keyword extraction methods, which typically focus on the frequency or importance of terms in the document using post-hoc analysis. Instead, we analyze the impact of each word on the spatial representation of a document in the projection space, using gradient tracking methods that are contextualized within the DR and embedding algorithms. Thus, our method is grounded in the semantics of the projection and the underlying embedding.  Therefore, our method more directly explains the semantics of the space and what words influenced the computation of the space. To demonstrate this, we apply our spatial word clouds to an alternative method for quantifying the impact of individual words - attention values in the BERT model. For each document, we extract the attention values for each word and input those into the spatial word clouds, rather than the gradients.

\cref{fig:case_study_attention} shows the spatial word clouds generated using the attention values. These are generated from the same projections as in \cref{sec:bbc_case_study} and are comparable to the gradient-based clouds in \cref{fig:case_study_bbc}. While the spatial word clouds using attention values identify some contextual information relevant to the embedding model, they fail to demonstrate the information impactful to the DR, presenting the same set of impactful words regardless of the DR algorithm.   In contrast, gradients capture the impacts of individual words through the entire pipeline - from the document words, through the embedding model and to the DR space.

\section{Discussion}

\textbf{Advantages Over Counterfactual Methods} Our approach introduces a more effective way of understanding text projections compared with the existing method presented by Bian et al., which relies on generating counterfactuals for each document\cite{bian2021semantic}. It does so by removing all instances of a word,
re-generating the embedding, and projecting it out-of-sample into the 2D space. While the resulting explanations are similar to ours, our system has three benefits over counterfactuals. First, our system calculates the impact of individual instances of words rather than the combined impact of all instances. This can be important as the contextual meaning of individual instances can vary depending on the context words around them. Furthermore, removing all instances of a word fundamentally changes the meaning of the document when generating the counterfactual embedding. Second, the counterfactual embedding is projected out-of-sample into the MDS space after the rest of the documents have been placed. Because MDS organizes points based on pairwise distances, this may not reflect the true placement of the counterfactual embedding. In contrast, our method calculates the gradients directly from the projection computation, without altering the meaning of the document. Thus, it more accurately reflects the impact of the underlying words on the projection. Third, counterfactuals are costly to compute due to combinatorial runs, whereas our method only requires 2 backwards passes per document.

\textbf{Generalizability} Our approach can be applied to a broader range of analytical pipelines. By adapting the algorithm to support automatic differentiation, our system is applicable across pipelines of projections, interactive analyses, and intermediate processing stages.

\textbf{Scalability} Our visualization techniques, such as spatial word clouds, enhance the system's scalability by efficiently aggregating datasets with a large number of words. By taking into account the spatial distribution of words with the associated documents in the projection space, our system ensures the visualization remains intuitive and insightful. However, we observed that some areas within the visualization become word-dense, resulting in some overlap. Therefore, developing improved visualization strategies is needed to handle such complexities in future work, especially for very large datasets.

\textbf{Performance} In terms of performance time, an advantage of our approach is that the gradients are computed as a by-product of the LLM and DR computation in Torch with autograd, requiring only $2n$ backwards passes, where $n$ is the number of documents. In our experience a backwards pass takes $\approx 1$ second. These could be computed in parallel, however Torch currently does not allow this. There is also a minor time cost for the tracking of the gradients with autograd. In our experience, enabling autograd increases algorithm execution time by approximately 12 percent.  The only additional execution is the spatial word-cloud algorithm, which has $O(n\log n)$
performance in average cases.

\section{Conclusion}
In this study, we present a method to enhance understanding of document projection through gradient analysis of text data. We designed a visualization system to assist users in interpreting the projection of documents, which includes  impact heatmaps, impactful words of individual documents and spatial word clouds of global word impacts. The three usage scenarios demonstrated how our visual system, supported by the gradient-based explanation, facilitates the identification and understanding of cluster patterns and relationships in the text data and enables comparison of different analysis settings. Additionally, our method can be extended to a variety of DR pipelines to illustrate the spatial semantics for different projection spaces.

\acknowledgments{
This material is based upon work supported by the National Science Foundation under Grant \# 2127309 to the Computing Research Association for the CIFellows 2021 Project.}

\bibliographystyle{abbrv-doi-hyperref}

\bibliography{template}

\begin{thebibliography}{10}

\bibitem{abadi2019tensorflow}
M.~Abadi, A.~Agarwal, P.~Barham, E.~Brevdo, Z.~Chen, C.~Citro, G.~S. Corrado, A.~Davis, J.~Dean, M.~Devin, et~al.
\newblock Tensorflow: Large-scale machine learning on heterogeneous systems (2015), software available from tensorflow. org.
\newblock {\em URL https://www. tensorflow. org}, 2019.

\bibitem{bian2021deepsi}
Y.~Bian and C.~North.
\newblock Deepsi: Interactive deep learning for semantic interaction.
\newblock In {\em 26th International Conference on Intelligent User Interfaces}, pp. 197--207, 2021.

\bibitem{bian2021semantic}
Y.~Bian, C.~North, E.~Krokos, and S.~Joseph.
\newblock Semantic explanation of interactive dimensionality reduction.
\newblock In {\em 2021 IEEE Visualization Conference (VIS)}, pp. 26--30. IEEE, 2021.

\bibitem{bradel2014multi}
L.~Bradel, C.~North, and L.~House.
\newblock Multi-model semantic interaction for text analytics.
\newblock In {\em 2014 IEEE Conference on Visual Analytics Science and Technology (VAST)}, pp. 163--172. IEEE, 2014.

\bibitem{bradel2015big}
L.~Bradel, N.~Wycoff, L.~House, and C.~North.
\newblock Big text visual analytics in sensemaking.
\newblock In {\em 2015 Big Data Visual Analytics (BDVA)}, pp. 1--8. IEEE, 2015.

\bibitem{burch2014radcloud}
M.~Burch, S.~Lohmann, F.~Beck, N.~Rodriguez, L.~Di~Silvestro, and D.~Weiskopf.
\newblock Radcloud: Visualizing multiple texts with merged word clouds.
\newblock In {\em 2014 18th International Conference on Information Visualisation}, pp. 108--113. IEEE, 2014.

\bibitem{cheng2020dece}
F.~Cheng, Y.~Ming, and H.~Qu.
\newblock Dece: Decision explorer with counterfactual explanations for machine learning models.
\newblock {\em IEEE Transactions on Visualization and Computer Graphics}, 27(2):1438--1447, 2020.

\bibitem{chi2015morphable}
M.-T. Chi, S.-S. Lin, S.-Y. Chen, C.-H. Lin, and T.-Y. Lee.
\newblock Morphable word clouds for time-varying text data visualization.
\newblock {\em IEEE transactions on visualization and computer graphics}, 21(12):1415--1426, 2015.

\bibitem{vcik2021explaining}
I.~{\v{C}}{\'\i}k, A.~D. Rasamoelina, M.~Mach, and P.~Sin{\v{c}}{\'a}k.
\newblock Explaining deep neural network using layer-wise relevance propagation and integrated gradients.
\newblock In {\em 2021 IEEE 19th world symposium on applied machine intelligence and informatics (SAMI)}, pp. 000381--000386. IEEE, 2021.

\bibitem{ciorna2023interact}
V.~Ciorna, G.~Melan{\c{c}}on, F.~Petry, and M.~Ghoniem.
\newblock Interact: A visual what-if analysis tool for virtual product design.
\newblock {\em Information Visualization}, p. 14738716231216030, 2023.

\bibitem{collobert2011torch7}
R.~Collobert, K.~Kavukcuoglu, and C.~Farabet.
\newblock Torch7: A matlab-like environment for machine learning.
\newblock In {\em BigLearn, NIPS workshop}, number CONF, 2011.

\bibitem{devlin2018bert}
J.~Devlin, M.-W. Chang, K.~Lee, and K.~Toutanova.
\newblock Bert: Pre-training of deep bidirectional transformers for language understanding.
\newblock {\em arXiv preprint arXiv:1810.04805}, 2018.

\bibitem{dowling2018sirius}
M.~Dowling, J.~Wenskovitch, J.~Fry, L.~House, and C.~North.
\newblock Sirius: Dual, symmetric, interactive dimension reductions.
\newblock {\em IEEE transactions on visualization and computer graphics}, 25(1):172--182, 2018.

\bibitem{dowling2019interactive}
M.~Dowling, N.~Wycoff, B.~Mayer, J.~Wenskovitch, L.~House, N.~Polys, C.~North, and P.~Hauck.
\newblock Interactive visual analytics for sensemaking with big text.
\newblock {\em Big Data Research}, 16:49--58, 2019.

\bibitem{duarte2014nmap}
F.~S. Duarte, F.~Sikansi, F.~M. Fatore, S.~G. Fadel, and F.~V. Paulovich.
\newblock Nmap: A novel neighborhood preservation space-filling algorithm.
\newblock {\em IEEE transactions on visualization and computer graphics}, 20(12):2063--2071, 2014.

\bibitem{endert2013typograph}
A.~Endert, R.~Burtner, N.~Cramer, R.~Perko, S.~Hampton, and K.~Cook.
\newblock Typograph: Multiscale spatial exploration of text documents.
\newblock In {\em 2013 IEEE International Conference on Big Data}, pp. 17--24. IEEE, 2013.

\bibitem{endert2012semantic}
A.~Endert, P.~Fiaux, and C.~North.
\newblock Semantic interaction for visual text analytics.
\newblock In {\em Proceedings of the SIGCHI conference on Human factors in computing systems}, pp. 473--482, 2012.

\bibitem{endert2011unifying}
A.~Endert, P.~Fiaux, C.~North, et~al.
\newblock Unifying the sensemaking loop with semantic interaction.
\newblock In {\em IEEE Workshop on Interactive Visual Text Analytics for Decision Making at VisWeek 2011}, 2011.

\bibitem{faust2018dimreader}
R.~Faust, D.~Glickenstein, and C.~Scheidegger.
\newblock Dimreader: Axis lines that explain non-linear projections.
\newblock {\em IEEE transactions on visualization and computer graphics}, 25(1):481--490, 2018.

\bibitem{gomez2020vice}
O.~Gomez, S.~Holter, J.~Yuan, and E.~Bertini.
\newblock Vice: Visual counterfactual explanations for machine learning models.
\newblock In {\em Proceedings of the 25th International Conference on Intelligent User Interfaces}, pp. 531--535, 2020.

\bibitem{gomez2021advice}
O.~Gomez, S.~Holter, J.~Yuan, and E.~Bertini.
\newblock Advice: Aggregated visual counterfactual explanations for machine learning model validation.
\newblock In {\em 2021 IEEE Visualization Conference (VIS)}, pp. 31--35. IEEE, 2021.

\bibitem{gorski2020towards}
L.~Gorski, S.~Ramakrishna, and J.~M. Nowosielski.
\newblock Towards grad-cam based explainability in a legal text processing pipeline.
\newblock {\em arXiv preprint arXiv:2012.09603}, 2020.

\bibitem{greene2006practical}
D.~Greene and P.~Cunningham.
\newblock Practical solutions to the problem of diagonal dominance in kernel document clustering.
\newblock In {\em Proceedings of the 23rd international conference on Machine learning}, pp. 377--384, 2006.

\bibitem{griewank2008evaluating}
A.~Griewank and A.~Walther.
\newblock {\em Evaluating derivatives: principles and techniques of algorithmic differentiation}.
\newblock SIAM, 2008.

\bibitem{hamal2023interpreting}
S.~Hamal.
\newblock {\em Interpreting Dimension Reductions through Gradient Visualization}.
\newblock PhD thesis, Virginia Tech, 2023.

\bibitem{hearst2019evaluation}
M.~A. Hearst, E.~Pedersen, L.~Patil, E.~Lee, P.~Laskowski, and S.~Franconeri.
\newblock An evaluation of semantically grouped word cloud designs.
\newblock {\em IEEE transactions on visualization and computer graphics}, 26(9):2748--2761, 2019.

\bibitem{john2018multicloud}
M.~John, E.~Marbach, S.~Lohmann, F.~Heimerl, and T.~Ertl.
\newblock Multicloud: Interactive word cloud visualization for multiple texts.
\newblock {\em Proceeding of Graphical Interface}, pp. 25--32, 2018.

\bibitem{kokhlikyan2020captum}
N.~Kokhlikyan, V.~Miglani, M.~Martin, E.~Wang, B.~Alsallakh, J.~Reynolds, A.~Melnikov, N.~Kliushkina, C.~Araya, S.~Yan, et~al.
\newblock Captum: A unified and generic model interpretability library for pytorch.
\newblock {\em arXiv preprint arXiv:2009.07896}, 2020.

\bibitem{liu2018bridging}
S.~Liu, X.~Wang, C.~Collins, W.~Dou, F.~Ouyang, M.~El-Assady, L.~Jiang, and D.~A. Keim.
\newblock Bridging text visualization and mining: A task-driven survey.
\newblock {\em IEEE transactions on visualization and computer graphics}, 25(7):2482--2504, 2018.

\bibitem{lohmann2015concentri}
S.~Lohmann, F.~Heimerl, F.~Bopp, M.~Burch, and T.~Ertl.
\newblock Concentri cloud: Word cloud visualization for multiple text documents.
\newblock In {\em 2015 19th International Conference on Information Visualisation}, pp. 114--120. IEEE, 2015.

\bibitem{lundberg2017unified}
S.~M. Lundberg and S.-I. Lee.
\newblock A unified approach to interpreting model predictions.
\newblock {\em Advances in neural information processing systems}, 30, 2017.

\bibitem{marcilio2021explaining}
W.~E. Marcilio-Jr and D.~M. Eler.
\newblock Explaining dimensionality reduction results using shapley values.
\newblock {\em Expert Systems with Applications}, 178:115020, 2021.

\bibitem{marshall1994viki}
C.~C. Marshall, F.~M. Shipman~III, and J.~H. Coombs.
\newblock Viki: Spatial hypertext supporting emergent structure.
\newblock In {\em Proceedings of the 1994 ACM European conference on Hypermedia technology}, pp. 13--23, 1994.

\bibitem{moujahid2022combining}
H.~Moujahid, B.~Cherradi, M.~Al-Sarem, L.~Bahatti, A.~B. A. M.~Y. Eljialy, A.~Alsaeedi, and F.~Saeed.
\newblock Combining cnn and grad-cam for covid-19 disease prediction and visual explanation.
\newblock {\em Intelligent Automation \& Soft Computing}, 32(2), 2022.

\bibitem{panwar2020deep}
H.~Panwar, P.~Gupta, M.~K. Siddiqui, R.~Morales-Menendez, P.~Bhardwaj, and V.~Singh.
\newblock A deep learning and grad-cam based color visualization approach for fast detection of covid-19 cases using chest x-ray and ct-scan images.
\newblock {\em Chaos, Solitons \& Fractals}, 140:110190, 2020.

\bibitem{paulovich2012semantic}
F.~V. Paulovich, F.~M. Toledo, G.~P. Telles, R.~Minghim, and L.~G. Nonato.
\newblock Semantic wordification of document collections.
\newblock In {\em Computer Graphics Forum}, vol.~31, pp. 1145--1153. Wiley Online Library, 2012.

\bibitem{pedregosa2011scikit}
F.~Pedregosa, G.~Varoquaux, A.~Gramfort, V.~Michel, B.~Thirion, O.~Grisel, M.~Blondel, P.~Prettenhofer, R.~Weiss, V.~Dubourg, et~al.
\newblock Scikit-learn: Machine learning in python.
\newblock {\em the Journal of machine Learning research}, 12:2825--2830, 2011.

\bibitem{ribeiro2016should}
M.~T. Ribeiro, S.~Singh, and C.~Guestrin.
\newblock " why should i trust you?" explaining the predictions of any classifier.
\newblock In {\em Proceedings of the 22nd ACM SIGKDD international conference on knowledge discovery and data mining}, pp. 1135--1144, 2016.

\bibitem{rudin2022interpretable}
C.~Rudin, C.~Chen, Z.~Chen, H.~Huang, L.~Semenova, and C.~Zhong.
\newblock Interpretable machine learning: Fundamental principles and 10 grand challenges.
\newblock {\em Statistic Surveys}, 16:1--85, 2022.

\bibitem{4577920}
C.~Seifert, B.~Kump, W.~Kienreich, G.~Granitzer, and M.~Granitzer.
\newblock On the beauty and usability of tag clouds.
\newblock In {\em 2008 12th International Conference Information Visualisation}, pp. 17--25, 2008. \href{https://doi.org/10.1109/IV.2008.89}
{doi: {{%
10\hspace{.1pt}\discretionary{.}{%
}{.}\hspace{.4pt}1109\discretionary{/}{%
}{/}IV\hspace{.1pt}\discretionary{.}{%
}{.}\hspace{.4pt}2008\hspace{.1pt}\discretionary{.}{%
}{.}\hspace{.4pt}89}}}


\bibitem{selvaraju2017grad}
R.~R. Selvaraju, M.~Cogswell, A.~Das, R.~Vedantam, D.~Parikh, and D.~Batra.
\newblock Grad-cam: Visual explanations from deep networks via gradient-based localization.
\newblock In {\em Proceedings of the IEEE international conference on computer vision}, pp. 618--626, 2017.

\bibitem{sundararajan2017axiomatic}
M.~Sundararajan, A.~Taly, and Q.~Yan.
\newblock Axiomatic attribution for deep networks.
\newblock In {\em International conference on machine learning}, pp. 3319--3328. PMLR, 2017.

\bibitem{maaten-tsne-python}
L.~van~der Maaten.
\newblock A {Python} implementation of {t-SNE}.
\newblock \url{https://lvdmaaten.github.io/tsne/}.

\bibitem{viegas2009participatory}
F.~B. Viegas, M.~Wattenberg, and J.~Feinberg.
\newblock Participatory visualization with wordle.
\newblock {\em IEEE transactions on visualization and computer graphics}, 15(6):1137--1144, 2009.

\bibitem{wang2024empirical}
A.~Z. Wang, D.~Borland, and D.~Gotz.
\newblock An empirical study of counterfactual visualization to support visual causal inference.
\newblock {\em Information Visualization}, p. 14738716241229437, 2024.

\bibitem{wang2020recloud}
J.~Wang, J.~Zhao, S.~Guo, C.~North, and N.~Ramakrishnan.
\newblock Recloud: semantics-based word cloud visualization of user reviews.
\newblock In {\em Graphics Interface 2014}, pp. 151--158. AK Peters/CRC Press, 2020.

\bibitem{wexler2019if}
J.~Wexler, M.~Pushkarna, T.~Bolukbasi, M.~Wattenberg, F.~Vi{\'e}gas, and J.~Wilson.
\newblock The what-if tool: Interactive probing of machine learning models.
\newblock {\em IEEE transactions on visualization and computer graphics}, 26(1):56--65, 2019.

\bibitem{xu2016semantic}
J.~Xu, Y.~Tao, and H.~Lin.
\newblock Semantic word cloud generation based on word embeddings.
\newblock In {\em 2016 IEEE Pacific Visualization Symposium (PacificVis)}, pp. 239--243. IEEE, 2016.

\bibitem{zang2022evnet}
Z.~Zang, S.~Cheng, L.~Lu, H.~Xia, L.~Li, Y.~Sun, Y.~Xu, L.~Shang, B.~Sun, and S.~Z. Li.
\newblock Evnet: An explainable deep network for dimension reduction.
\newblock {\em IEEE Transactions on Visualization and Computer Graphics}, 2022.

\end{thebibliography}

\appendix 

\end{document}